\documentclass{article}
\usepackage{times,geometry,enumerate}
\usepackage[dvips]{graphicx}
\usepackage{subfigure}
\usepackage{hyperref}

\providecommand{\keywords}[1]{\textbf{Keywords:} #1}
\begin{document}
\author{Cristina Blaga\\Faculty of Mathematics and Computer Science \\Babe\c{s}-Bolyai University, Cluj-Napoca, Romania}
\title{Circular orbits, Lyapunov stability and Manev-type forces}
\date{}
\maketitle
\begin{abstract}
In this article we study the stability in the sense of Lyapunov of the circular orbits in the generalized Manev two bodies problem. First, we explore the existence of the circular orbits and determine their radius. Then, using the first integrals of motion we build a positive definite function, known as a Lyapunov function. It's existence proves that the circular orbit is stable in the sense of Lyapunov. In the end, we consider several real systems of two bodies and compare the characteristics of the circular orbits in Newtonian and modified Manev gravitational field, arguing about our possibilities to observe the differences between the motion in these two fields.
\end{abstract}


\section{Introduction}

In the third decade of the XX century, Georgi Manev proposed a new gravitational law instead the Newtonian gravitational law (\cite{m25}, \cite{m29}, \cite{m30}). According to it, on a particle of mass $m_2$, moving in the static gravitational field produced by the mass $m_1$ ($m_1>m_2$) acts the force  
\begin{equation}\label{fM}
	F_M(r)=-\frac{G m_1 m_2}{r^2} \left( 1 + \frac{3 G (m_1+m_2)}{c^2 r} \right)
\end{equation}
where $r$ is the distance from $m_2$ to the center of mass of $m_1$. The difference between the Manev force~(\ref{fM}) and the Newtonian force is the term inverse proportional to $r^3$. The additional term, allow us to explain phenomena observed in the solar system, like Mercury's perihelion precession, unexplained using the Newtonian gravitational force.

Sir Isaac Newton used a force similar to~(\ref{fM}) in order to explain the motion of the apsidal line of the Moon. In his book \emph{Philosophiae Naturalis Principia Mathematica}, he considered an additional term of the form $\mu/r^3$ (where $\mu>0$ and $r$ the distance between the particles) to the Newtonian gravitational force and proved that under such force the particle describes a precessional ellipse (see for example \cite{n99}, Book I, Section IX, Proposition XLIV, Theorem XIV).  

Manev nonrelativistic gravitational law appears almost at the same time with Einstein's general relativity. After the later theory was confirmed by observations, during the total solar eclipse from 1919, Manev's model was forgotten. It was rediscovered by \cite{d93} and since then used to explain various phenomena from astronomy like: free gravitational collapse \cite{u95}, stellar dynamics \cite{b97}, gravitational redshift \cite{u98} or the advance of Mercury's perihelion \cite{u99}. There are applications of Manev's model in other branches of sciences, like: chemistry, biology, \emph{a.o.} A comprehensive list of them could be found in \cite{hm09}. 

In his work, \cite{u99} emphasized that the Manev force~(\ref{fM}) explains the perihelion advance of planet Mercury qualitatively, but not quantitatively. The observational results are fully explained using a Manev-type force, if the coefficient of the term inverse proportional with $r^3$ is doubled. Such a modified Manev force was used by \cite{khg13}. The authors considered the motion of a satellite in low earth orbit, in a modified Manev gravitational field, under the action of a drag force proportional with the square of the velocity. 

In the frame of the inverse problem of dynamics, \cite{bl05} proved that a two-parametric family of preccesing conics is produced by a conservative central force of the form
\begin{equation}\label{fMg}
	F(r)=-\frac{\alpha}{r^2} - \frac{\beta}{r^3}
\end{equation}
with $\alpha$ and $\beta$ two real constants. The force~(\ref{fMg}) is known as \emph{Manev-type force} or \emph{generalized Manev force}. 

We consider here the stability in the sense of Lyapunov of circular orbits in the generalized Manev two bodies problem. The stability of the circular orbits in the classical Manev gravitational field was discussed in \cite{b15}. There we proved that the circular orbit in the two bodies problem under the action of a Manev central force~(\ref{fM}) is stable in the sense of Lyapunov. This means that if the motion of a particle describing a circular orbit is perturbed, after a while it will come back to the initial orbit.

We considered the circular orbits, because they play an important role in space dynamics. At the beginning of the space era, the orbiters of the Moon, Venus or Mercury were designed to describe segments of ellipses with small eccentricities. In that case, the near circular orbits were preferred, because if the orbiter would have moved on an elongated orbit, it could have stepped out from the sphere of influence of the surveyed body \cite{r04}. Even now, the orbiters surveying the small bodies from the solar system, like dwarf planets, asteroids or comet nuclei revolve around the body on near circular orbits, to avoid the exit from the sphere of influence of the body. After 2000, in the vicinity of Earth were launched satellites in low, near circular polar orbits to make precise measurements of Earth's gravity field (GRACE - twin satellites), Earth's magnetic field (SWARM - a constelation of three satellites) or both gravity and magnetic field (CHAMP mission between 2000 and 2010). 

In this article we study the stability of the circular orbits in the sense of Lyapunov, in the two bodies problem, under the action of a central Manev-type force~(\ref{fMg}). The equations of motion and the radius of the circular orbit are obtained in section two. The stability in the sense of Lyapunov is proved in the third section. The Lyapunov function is build upon the first integrals of motion. In section $4$, we consider real two-body systems, in which a small body describes an almost circular orbit around the parent body. For them, we compute the differences between the radius and period of the motion in the Newtonian and modified Manev field. We argue about our capacity to measure the obtained differences. The concluding remarks are gathered in the last section.   

\section{The generalized Manev two bodies problem}

Through \emph{generalized Manev two bodies problem} we understand the motion of two bodies of masses $m_1$ and $m_2$ ($m_1>m_2$), due to a central Manev-type force (\ref{fMg}), acting on the line joining the centers of mass of the bodies. This is a generalization of the classical Manev two bodies problem studied in~\cite{b15}.

\subsection{The equations of motion}

Consider two bodies of masses $m_1$ and $m_2$ ($m_1>m_2$), under the action of a central Manev-type force~(\ref{fMg}), assuming that the distance between the bodies equals the distance between their centers of mass and denoting their position vectors with $\vec{r_1}$ and $\vec{r_2}$, the equations of motion are
\begin{eqnarray}
	m_1 \ddot{\vec{r}}_1 &=& -F(r) \, \frac{\vec{r}}{r} \label{m1}\\
	m_2 \ddot{\vec{r}}_2 &=& F(r) \, \frac{\vec{r}}{r} \label{m2}
\end{eqnarray}
where $F(r)$ is the Manev-type force~(\ref{fMg}) and $\vec{r}=\vec{r}_2 - \vec{r}_1$ is the position vector of the mass $m_2$ relative to the mass $m_1$. From the equations of absolute motion~(\ref{m1}) and~(\ref{m2}), we obtain the equation of the relative motion of $m_2$ in respect with $m_1$
\begin{equation}
	\mu \ddot{\vec{r}}=F(r) \frac{\vec{r}}{r} \label{mr} \,,
\end{equation} 
where $\mu = m_1 m_2/(m_1 + m_2)$ is the reduced mass of the bodies.

Like Manev force~(\ref{fM}), the generalized Manev force~(\ref{fMg}) could be derived from a scalar function $V(r)$: $F(r)= -d V(r)/d r$. The function $V(r)$ is called \emph{Manev-type potential} or \emph{generalized Manev potential} and is given by  
\begin{equation}\label{VMg}
	V(r)=-\frac{\alpha}{r} - \frac{\beta}{2 r^2}\,,
\end{equation}
where $\alpha$ and $\beta$ are two real positive constants. It appears in the energy integral. To obtain this first integral of motion, we perform the scalar multiplication of~(\ref{mr}) with the relative velocity of $m_2$ with respect to $m_1$ ($\vec{v}=\dot{\vec{r}}$) and get 
\begin{equation}\label{ie}
	\frac{\mu}{2} v^2 + V(r) = h, 
\end{equation}
where $v$ is the length of velocity and $h$ the total energy. 

The angular momentum integral is obtained through vectorial multiplication of~(\ref{mr}) with the position vector $\vec{r}$. We get   
\begin{equation}\label{iav}
	\vec{r} \times \mu \vec{\dot{r}} = \vec{C},
\end{equation}
where $\vec{C}$ is a constant vector, normal to the plane of motion. It means that, the motion is planar, like in the Newtonian and classical Manev case. The motion take place in the plane given by the initial position of $m_1$, the initial position vector of $m_2$ relative to $m_1$ and initial velocity of $m_2$ relative to $m_1$.  

Further, we will study the motion in the orbital plane, using polar coordinates ($r$,$\varphi$). The projections of equation of relative motion~(\ref{mr}) on the radial and polar axis are 
\begin{eqnarray}
	\ddot{r}-r {\dot{\varphi}}^2 &=& - \frac{\alpha}{\mu r^2} \left( 1 + \frac{\beta}{\alpha r} \right) \label{mrer}\\
	\frac{1}{r} \frac{d}{d t} (r^2 \dot{\varphi}) &=& 0, \label{ia}
\end{eqnarray}      
with $\mu$ the reduced mass and $\alpha$, $\beta$ two positive constants. 

In polar coordinates, the angular momentum integral~(\ref{iav}) becomes $|\vec{C}| = \mu r^2 \dot{\varphi}$. In other words, the relation~(\ref{ia}) is equivalent with the integral of angular momentum~(\ref{iav}).   

\subsection{The existence of circular orbits}

The radius of the circular orbits is obtained from the equations~(\ref{mrer}) and~(\ref{ia}). The circular motion is given by $r=r_0=\mbox{constant}$. Thus $\dot{r}=\ddot{r}=0$ and from~(\ref{ia}), we get $\dot{\varphi}=C/r_0^2$. After some direct calculation, from~(\ref{mrer}) we obtain
\begin{equation}\label{r0g}
	r_0=\frac{\mu C^2}{\alpha}\left( 1- \frac{\beta}{\mu C^2} \right)
\end{equation}
and conclude that in the generalized Manev two bodies problem exists one circular orbit if $C^2>\beta/\mu$. 

The relation~(\ref{r0g}) generalize the formula of the radius of the circular orbit in classical Manev two bodies problem and Kepler problem. If $\alpha=G m_1 m_2$ and $\beta=3 G^2 m_1 m_2 (m_1+m_2)/c^2$, where $G$ is the gravitational parameter and $c$ the speed of light, denoting $\gamma=G(m_1+m_2)$, the relation~(\ref{r0g}) becomes
\begin{equation}\label{r0}
	r_0=\frac{C^2}{\gamma}\left( 1- \frac{3 \gamma^2}{C^2 c^2} \right)\,,
\end{equation} 
the formula of the radius of the circular orbit in classical Manev two bodies problem \cite{b15}. If $\alpha=G m_1 m_2$ and $\beta=0$, then~(\ref{r0g}) becomes $r_{0}=C^2/\gamma$, the formula of the radius of the circular orbit in the classical Newtonian two bodies problem \cite{g80}.

\section{The stability of the circular orbit}

We analyze the stability in the sense of Lyapunov of the circular orbit in the generalized Manev two bodies problem like the stability of the circular orbit in the classical Manev two bodies problem \cite{b15}.

We use the spherical coordinates ($r$, $\theta$, $\varphi$), with $r$ - the radial coordinate, $\theta$ - the latitude and $\varphi$ - the longitude and their time derivatives $\dot{r}$, $\dot{\theta}$, $\dot{\varphi}$ and identify the unperturbed motion with the motion on the circular orbit. In this 6-dimensional space, the motion on the circular orbit is specified by the point $(r_0,0,\varphi_0,0,0,\dot{\varphi_0})$, denoted with $\mathbf{0}$, where $r_0$ is the radius of the circular orbit~(\ref{r0g}), $\theta_0=0$ - we assume that the motion take place in the equatorial plane, $\varphi_0$ is the initial longitude and $\dot{\varphi_0}=C/r_0^2$ - according to the angular momentum integral. This point is an equilibrium point for the equations of motion~(\ref{mrer}) and~(\ref{ia}). 

We introduce the Lagrangian 
\begin{equation}\label{L}
	L=T-V(r) 
\end{equation}
where $V(r)$ is the generalized Manev potential from~(\ref{VMg}) and the kinetic energy $T$ is given by 
\begin{equation}
	T=\frac{\mu}{2}\left( \dot{r}^2 + r^2 \dot{\theta}^2 + r^2 \cos^2 \theta  \dot{\varphi}^2 \right) \,.
\end{equation} 

The longitude $\varphi$ is a cyclic coordinate for the Lagrangian~(\ref{L}), therefore
\begin{equation}\label{F2}
	r^2 \cos ^2 \theta \dot{\varphi} = b \,,
\end{equation}   
with $b$ a constant. It follows that only five, from the coordinates are independent. The perturbed motion is given by $x_i$, $i=\overline{1,5}$, 
\begin{equation}
	r = r_0 +x_1 ,\quad \dot{r} = x_2 ,\quad \theta = x_3 ,\quad \dot{\theta} = x_4 ,\quad \dot{\varphi}=\dot{\varphi_0} + x_5 ,
\end{equation}
with $\dot{\varphi_0}=C/r_0^2$ and $r_0$ from~(\ref{r0g}).

The theorem of stability in sense of Lyapunov states that an equilibrium point of the differential equations is stable if there exists a positive definite function $\mathcal{F}$, whose derivative $\dot{\mathcal{F}}$ for the given differential equations is a negative semi-definite function or identically zero in a neighborhood of the equilibrium point (Theorem 25.1, page 102, \cite{h67}). The function $\mathcal{F}$ is known as Lyapunov function. There is no general algorithm to build it. Usually, the first integrals of motion are the bricks used to build this function. 

In the case of the generalized Manev two bodies problem, for perturbed motion, the energy integral~(\ref{ie}), after the division by $\mu/2$, becomes
\begin{equation}\label{F1p}
F_1(\mathbf{x}) = x_2^2+(r_0+x_1)^2 x_4^2 +(r_0+x_1)^2 (\dot{\varphi}+x_5)^2\cdot \cos^2 x_3 - \frac{2 \alpha}{\mu(r_0+x_1)} - \frac{\beta}{\mu (r_0+x_1)^2}.
\end{equation} 
The angular momentum integral~(\ref{F2}) is
\begin{equation}\label{F2p}
	F_2(\mathbf{x})=(r_0+x_1)^2 (\dot{\varphi}+x_5) \cos^2 x_3 \,.
\end{equation} 
These functions are not positively defined in a neighborhood of the unperturbed motion, reason why we look for a Lyapunov function of the form
\begin{equation}\label{F}
	\mathcal{F} (\mathbf{x}) = F_1(\mathbf{x})-F_1(\mathbf{0})+\lambda [F_2(\mathbf{x})-F_2(\mathbf{0})] + \nu [F_2^2(\mathbf{x})-F_2^2(\mathbf{0})], 
\end{equation}
where $\lambda$, $\nu$ are two real constants. We will try to find out, if there are two real positive numbers $\lambda$ and $\nu$, so that $\mathcal{F} (\mathbf{x})$ is a Lyapunov function. For that, we expand in Taylor series the terms from~(\ref{F}), keeping only the terms up to the second order. Having in mind that 
\begin{equation}\label{pp0}
	r_0 \dot{\varphi_0}^2= \frac{\alpha}{\mu r_0^2} \left(1+\frac{\beta}{\alpha r_0} \right),
\end{equation}    
we obtain
\begin{equation}\label{F1x}
	F_1(\mathbf{x})-F_1(\mathbf{0}) = 4 r_0 \dot{\varphi}_0^2 x_1 + 2 r_0^2 \dot{\varphi}_0 x_5 - \left( \dot{\varphi}_0^2 + \frac{\beta}{\mu r_0 ^4}\right) x_1^2 + x_2^2 - r_0^2 \dot{\varphi}_0^2 x_3^2 + r_0 \left( x_4^2 + x_5^2 \right) + 4 r_0 \dot{\varphi}_0 x_1 x_5, 
\end{equation}   
\begin{equation}\label{F2x}
	F_2(\mathbf{x})-F_2(\mathbf{0}) = 2 r_0 \dot{\varphi}_0 x_1 + r_0^2 x_5 + \dot{\varphi}_0 x_1^2 - r_0^2 \dot{\varphi}_0 x_3^2 + 2 r_0 x_1 x_5\, ,
\end{equation}
\begin{equation}\label{F22x}
	F_2(\mathbf{x})^2-F_2(\mathbf{0})^2 = 4 r_0^3 \dot{\varphi}_0^2 x_1 + 2 r_0^4 \dot{\varphi}_0 x_5 + 6 r_0^2 \dot{\varphi}_0^2 x_1^2 - 2 r_0^4 \dot{\varphi}_0^2 x_3^2 + 8 r_0^3 \dot{\varphi}_0 x_1 x_5 + r_0^4 x_5^2\,.
\end{equation}
Analyzing the expression of $\mathcal{F} (\mathbf{x})$ obtained after the substitution of~(\ref{F1x}),~(\ref{F2x}) and~(\ref{F22x}) in~(\ref{F}), we get that its derivative is identically zero in a neighborhood of the unperturbed motion if and only if 
\begin{equation}\label{l}
	\lambda = - 2 \dot{\varphi}_0 \left( 1 + r_0^2 \nu \right) \,.
\end{equation}
If~(\ref{l}) holds, we can write $\mathcal{F} (\mathbf{x})$ as a sum of two functions
\begin{equation}
	\mathcal{F} (\mathbf{x}) = \mathcal{F}_1 (x_2,x_3,x_4) + \mathcal{F}_2 (x_1,x_5)\, ,
\end{equation}
where
\begin{equation}
	\mathcal{F}_1 (x_2,x_3,x_4) = x_2^2 + r_0^2 \dot{\varphi}^2 x_3^2 + r_0^2 x_4^2\, ,
\end{equation}
\begin{equation}
	\mathcal{F}_2 (x_1,x_5) = c_{11} x_1^2 + 2 c_{12} x_1 x_5 + c_{22} x_5\, ,
\end{equation}
with 
\begin{equation}\label{c}
	c_{11}=4 \dot{\varphi}_0^2 \left(r_0^2 \nu - 1 \right) + \frac{\alpha}{\mu r_0^3}, \quad c_{22}= r_0^2 \left( 1 + \nu r_0^2 \right), 
	\quad c_{12}=c_{21}=2 r_0^3 \dot{\varphi}_0 \nu. 
\end{equation}
We notice that $\mathcal{F}_1$ is a positively defined function in $x_2$, $x_3$ and $x_4$. The other part of $\mathcal{F}$, the function $\mathcal{F}_2$ is positively defined in $x_1$ and $x_5$ if and only if all the principal minors of the quadratic form are positive, \emph{i.e.} 
\begin{equation}\label{pd}
	c_{11}>0 \quad \mbox{and} \quad d=c_{11} c_{22}-c_{12}^2>0\,.
\end{equation} 
Substituting~(\ref{c}) in~(\ref{pd}) we get
\begin{equation}
	d=\frac{\alpha}{\mu r_0} (1+\nu r_0^2) -4 r_0^2 \dot{\varphi}_0^2\,.
\end{equation}
The conditions~(\ref{pd}) are fulfilled if  
\begin{equation}\label{nu}
	\nu> \max \left\lbrace \nu_1, \nu_2  \right\rbrace 
\end{equation}
where
\begin{equation}
	\nu_1= \frac{3 \alpha r_0 +4 \beta}{4 \mu r_0^6 \dot{\varphi}_0^2} \quad \mbox{and} \quad \nu_2= \frac{3 \alpha r_0 + 4 \beta}{\alpha r_0^3}\,.
\end{equation}
We notice that $\nu_1>0$, $\nu_2>0$ and $\nu_2/\nu_1=4 \left(1+\beta r_0/\alpha \right) >4$, because $r_0$, $\alpha$ and $\beta$ are positive real numbers and using~(\ref{pp0}). Therefore, $\nu_2>4 \nu_1$ and the condition~(\ref{nu}) reduces to $\nu>\nu_2$. Thus, if $\nu$ satisfies the previous inequality, the function $\mathcal{F}_2 (x_1,x_5)$ is positively defined. Therefore, the function $\mathcal{F} (\mathbf{x})$ is positively defined and the circular orbit~(\ref{r0g}) is stable in the sense of Lyapunov. 

This is a generalization of the Lyapunov stability of circular orbits in the classical Manev two bodies problem \cite{b15} and Kepler problem \cite{d76}. It means that a small perturbation of $m_2$ from the circular orbit~(\ref{r0}), will never departure $m_2$ from the original orbit, in time $m_2$ will come back to its initial orbit~\cite{w27}. 

\section{Numerical results and discussion}

Based on the fact that the circular orbit in the generalized Manev two bodies problem is stable in the sense of Lyapunov, we consider several real two-body systems in which a small body describes an almost circular orbit around a massive body. We assume that upon the small body acts of a modified Manev force
\begin{equation}\label{fMm}
	F_{m}(r)= -\frac{G m_1 m_2}{r^2} \left( 1 + \frac{6 G(m_1+m_2)}{c^2 r} \right)
\end{equation}  
and compare the features of its motion on a circular orbit in Newtonian and modified Manev field. We choose this force, because using it we can explain qualitatively \emph{and} quantitatively the observed perihelion advance of Mercury (Ureche 1999).  

The radius of the circular orbit~(\ref{r0g}) under the action of~(\ref{fMm}) is
\begin{equation}\label{r0m}
	r_{0m}=\frac{C^2}{G(m_1+m_2)}\left( 1- \frac{6 G^2(m_1+m_2)^2}{c^2 C^2} \right)\,,
\end{equation}
and for a given value of the constant of the angular momentum $C$, the difference between radius of circular orbit in the Newtonian ($r_{0N}$) and modified Manev force field ($r_{0m}$) is
\begin{equation}\label{dr0}
	\Delta r_0=r_{0N}-r_{0m} = \frac{6 G(m_1+m_2)}{c^2} \,.
\end{equation}  

During the motion $C=r_0^2 \dot{\varphi}_0=\mbox{constant}$, therefore the circular motion is uniform in both fields (Manev and Newtonian). For a given angular momentum $C$, $r_{0m} < r_{0N}$, therefore $\dot{\varphi}_{0m} > \dot{\varphi}_{0N}$ and between the periods of motion there is the relation $T_{0m}=2 \pi/\dot{\varphi}_{0m} < T_{0N} = 2 \pi/\dot{\varphi}_{0N}$. For a given $C$, the difference between the periods of circular motion in the Newtonian and Manev gravitational field is
\begin{equation}\label{dT0}
	\Delta T_0=T_{0N}-T_{0m} = \frac{24 \pi C}{c^2}\left( 1- \frac{3 G^2(m_1+m_2)^2}{c^2 C^2} \right) \,.
\end{equation}     

Further, we compute the differences between radius and periods of circular motion in modified Manev and Newtonian gravitational field for two-body systems like: Earth and an artificial satellite, a planet from solar system and its moon and a star and a planet orbiting and discuss if we are able to measure these quantities. 

For numerical investigation we chose two Earth artificial satellites with almost circular orbits: LARES and GPS IIF-9. Their mean motion, semimajor axis and eccentricities are given in Table 1. 

\begin{table}
	\centering
	\caption{The mean motion, eccentricity and semimajor axis of the satellites LARES at epoch 2015 June 8, 21.4 UT and GPS BIIF-9 at 2015 June 9, 12.9 UT. Source: http://www.celestrak.com}\label{tab1}
	{\footnotesize
		\begin{tabular}{|c|c|c|}
			\hline Satellite&LARES&GPS BIIF-9
			\\
			\hline
			Mean motion (rev/d)&12.549&2.0058\\ \hline
			Eccentricity&0.0011222&0.0001283\\ \hline
			Semimajor axis (km)&7822.294&26560.05\\ 
			\hline
		\end{tabular}
	}
\end{table}

LARES (LAser RElativity Satellite) is a low Earth orbiter (LEO), laser-ranged satellite, designed to study Lense-Thirring effect in Earth vicinity. GPS IIF-9 is the ninth satellite from the Block IIF GPS navigation system, operating in medium Earth orbit (MEO). We have selected these satellites because they have almost circular orbits and their orbit could be reconstructed with centimeter accuracy. The method used for a low Earth orbiter is described in \cite{c14} and for a GPS satellite in \cite{b05}.

LARES is a sphere of tungsten alloy of 400 kg mass and GPS BIIF-9 had a launch mass of 1630 kg. The masses of the satellites are much smaller than Earth's mass ($M_{\oplus}=5.974 \times 10^{24}$ kg), therefore in both cases, the difference between the radius of circular motion in the Newtonian and Manev field, $\Delta r_0$ from~(\ref{dr0}), is about $2.7$ $cm$ and the difference between the periods of motion, $\Delta T_0$ from~(\ref{dT0}), are about $47$ $\mu s$ for LARES and $86$ $\mu s$ for GPS BII-F 9. For these satellites the orbit could be determined with centimeter accuracy, therefore we conclude that if their orbits were circular, we would have been able to measure the effect on their orbit of the additional term proportional with $r^{-3}$ from Manev force. But, their orbits are near circular and the variation of $r$ due to orbit's eccentricity is greater than the distance $\Delta r_0$ computed with formula~(\ref{dr0}).  

From the solar system we have selected two planetary satellites: Deimos - the second Mars satellite and Ganymede - the largest Jupiter satellite, both having almost circular orbits. Their orbital elements are given in Table 2.     

\begin{table}
	\centering
	\caption{The masses of Deimos and Ganymede and semimajor axis and eccentricity of Deimos at 1950 Jan. 1.00 TT and Ganymede at 1997 Jan. 16.00 TT. Source: http://ssd.jpl.nasa.gov/?sat\_elem}\label{tab2}
	{\footnotesize
		\begin{tabular}{|c|c|c|}
			\hline Satellite&Deimos&Ganymede\\ \hline
			Mass (kg)&$1.8\cdot 10^{15}$&$1.5\cdot 10^{23}$\\ \hline
			Eccentricity&0.0002&0.00013\\ \hline
			Semimajor axis (km)&23458&1070400\\ \hline
		\end{tabular}
	}
\end{table}

Replacing the orbital elements from Table 2 in the formulae~(\ref{dr0}) and~(\ref{dT0}), we got for the system Mars-Deimos $\Delta r_{0m}=2.9$ $mm$ and $\Delta T_0=26.9$ $\mu s$. These quantities are too small to be measured from Earth. When Mars is at opposition, from Earth the difference $\Delta r_{0m}$ is seen under an angle of $7.75 \cdot 10^{-9}$ arcseconds. 

For Jupiter and Ganymede, the differences in radius and periods of motion are $\Delta r_{0m}=8.44$ $m$ and $\Delta T_0=9.75$ $m s$. Although the differences are greater than in the case of Deimos, the maximum value of the angular distance corresponding to $\Delta r_{0m}$ is $2.8 \cdot 10^{-6}$ arcseconds. We need a powerful telescope to measure this small angle and as in the case of Earth's artificial satellites the variation of the distance planet-moon due to orbits eccentricity is greater than $\Delta r_{0m}$ from~(\ref{dr0}).

From the exoplanets database, \emph{The Extrasolar Planets Encyclopaedia}, available at http://exoplanet.eu/, we chose two planets with almost circular orbits: 55 Cnc f and HD 177830 b. The mass of the planets, their semimajor axis and eccentricity are given in Table 3. 

\begin{table}
	\centering
	\caption{The mass, semimajor axis and eccentricity of the exoplanets 55 Cnc f and HD 177830 b respectively. The mass is given in relative jovian masses (the mass of Jupiter in Earth masses is $M_J=317.8 M_{\oplus}$) and the semimajor axis in astronomical units (1 AU=$1.496 \cdot 10^{11} m$). Source: http://exoplanet.eu/}\label{tab3}
	{\footnotesize
		\begin{tabular}{|c|c|c|}
			\hline Planet&55 Cnc f&HD 177830 b\\ \hline
			Mass (in jovian masses)&$0.144$&$1.49$\\ \hline
			Eccentricity&0.0002&0.0009\\ \hline
			Semimajor axis (in AU)&$0.781$&$1.2218$\\ \hline
		\end{tabular}
	}
\end{table}

In Table 4 we give the relative masses of the stars (55 Cnc and HD 177830), the distance and the number of the planets orbiting them.   

\begin{table}
	\centering
	\caption{ The masses, distances and number of plantes orbiting the stars 55 Cnc and HD 177830, respectively. The masses are given in relative solar masses. The solar mass $(M_{\odot}=2 \cdot 10^{30} kg)$ and the distance in parsecs (1 pc=$206265 AU$). Source: http://exoplanet.eu/}\label{tab4}
	{\footnotesize
		\begin{tabular}{|c|c|c|}
			\hline Star&55 Cnc&HD 177830 \\ \hline
			Mass (in solar masses)&$0.905$&$1.47$\\ \hline
			Distance (in pc)&$12.34$&$59.0$\\ \hline
			Number of planets&5&2\\ \hline
		\end{tabular}
	}
\end{table}

Using the parameters of the stars and exoplanets we have computed the differences $\Delta r_0$ from~(\ref{dr0}) and $\Delta T_0$ from~(\ref{dT0}). For 55 Cnc f we have obtained that $\Delta r_{0m}=8 \, km$, $\Delta T_0=3.14 \, s$ and for HD 177830 b: $\Delta r_{0m}= 14 \, km$ and $\Delta T_0=5 \,s$. The two radial distances are seen from Earth under an angle of $4.3 \cdot 10^{-9}$ arcseconds, respectively $1.5 \cdot 10^{-9}$ arcseconds. These differences are to small to be determined with today observational techniques. 

\section{Conclusions}

Our task was to investigate the stability in sense of Lyapunov of circular orbits in the generalized Manev two-body problem. The motion is due to a central force of the form $F(r)=-\alpha/r^2-\beta/r^3$, with $\alpha$ and $\beta$ two positive real numbers. We compare our results with those from the classical Manev force, for which $\alpha=G m_1 m_2$ and $\beta=3 G^2 m_1 m_2 (m_1+m_2)/c^2$ and the Newtonian force, $\alpha=G m_1 m_2$ and $\beta=0$. 

For two given bodies of masses $m_1$ and $m_2$ ($m_1>m_2$) under the action of the central Manev-type force, the radius of the circular orbit of $m_2$ depends on $C$ - the angular momentum constant. Based on the first integral of angular momentum~(\ref{iav}) for the circular orbit, we got that the angular velocity $\dot{\varphi}_{0}$ is constant, circular motion is uniform, like in the classical Manev \cite{b15} or in the Newtonian two bodies problem \cite{g80}. 

The Lyapunov function was built using the first integrals of motion. The relation between $\nu$ and $\lambda$ has the same analytical expression $\lambda = - 2 \dot{\varphi}_0 \left( 1 + r_0^2 \nu \right)$ in the generalized and classical Manev \cite{b15}, respectively Newtonian two bodies problem \cite{d76}, but we note that, for a given value of the angular momentum $C$ and $\nu$, we get different values for $\lambda$, because $r_0$ -- the radius of the circular orbit, has different values in those three cases. 

The condition for the Lyapunov stability of the circular orbit in the Manev-type potential $$\nu > \frac{3 \alpha r_0 + 4 \beta}{\alpha r_0^3}$$ is reduced to the corresponding conditions from the two-body problem in the classical Manev \cite{b15} or in the Newtonian two bodies problem \cite{g80}. Therefore, we conclude that these results generalize the results from the classical Manev \cite{b15} and Newtonian two bodies problem \cite{d76}. The general form of the Manev force enable us to explain different dynamical phenomena observed in the solar system, unexplained in the Newtonian gravitational field, preserving the Lyapunov stability of the circular orbit.

The numerical exploration from the previous section revealed that for the selected two-body systems, for a given angular momentum constant $C$, determined of the initial conditions ($C=|\vec{r_0} \times \vec{v_0}|/\mu$), the differences between the radius of the circular orbit in the Newtonian and Manev field are small. This means that we need very accurate observations to decide whether the body $m_2$ describes a circular orbit around $m_1$ in a Manev-type or in a Newtonian field. But, we have proved that the circular motion is stable in sense of Lyapunov in a Manev-type gravitational field, like in the Newtonian field. 

\vspace{0.5cm}
\emph{Acknowledgements}. This paper was presented at the conference "Vistas in Astronomy, Astrophysics and Space Sciences", 30 - 31 May 2016, Cluj-Napoca, Romania.

\end{document}